\documentclass[12pt]{article}
\usepackage{lineno}
\usepackage{booktabs}
\usepackage{graphicx}
\usepackage{upgreek}
\usepackage{placeins}
\usepackage{comment}
\usepackage{hyperref}
\usepackage{subcaption}
\usepackage{multirow}

\newcommand\pubdate{\today}

\def\Title#1{\begin{center} {\Large #1 } \end{center}}
\def\Author#1{\begin{center}{ \sc #1} \end{center}}
\def\Address#1{\begin{center}{ \it #1} \end{center}}

\newcommand\pubblock{\rightline{\begin{tabular}{l}  \\ 
         \pubdate  \end{tabular}}}
\newenvironment{Abstract}{\begin{quotation}  }{\end{quotation}}
\newenvironment{Presented}{\begin{quotation} \begin{center}
             PRESENTED AT\end{center}\bigskip
      \begin{center}\begin{large}}{\end{large}\end{center} \end{quotation}}

\begin{document}
\begin{titlepage}
 \pubblock
\vfill
\Title{Measurement of $\Lambda$ hyperon spin-spin correlations in p+p collisions by the STAR experiment}
\vfill
\Author{ Jan Vanek, for the STAR collaboration }
\Address{Brookhaven National Laboratory}
\vfill
\begin{Abstract}
Polarization of $\Lambda$ hyperons has been observed in various collision systems over a  wide range of collision energies over the last 50 years since its discovery in Fermilab in the 70's. The existing experimental and theoretical techniques were not able to provide a conclusive answer about the origin of the polarization. In these proceedings, we discuss the possibility to use a new experimental method which utilizes measurement of $\Lambda\bar{\Lambda}$, $\Lambda\Lambda$, and $\bar{\Lambda}\bar{\Lambda}$ pair spin-spin correlations. With this new approach, it should be possible to distinguish if the polarization originates from early stage effects, such as initial state parton spin correlation, or if it is a final state effect originating from hadronization. Furthermore we, study the feasibility to perform this measurement in $p+p$ collisions at $\sqrt{s} = 200$~GeV collected by STAR in 2012 which should provide sufficient statistics of  $\Lambda\bar{\Lambda}$, $\Lambda\Lambda$, and $\bar{\Lambda}\bar{\Lambda}$ pairs to perform this measurement.
\end{Abstract}
\vfill
\begin{Presented}
DIS2023: XXX International Workshop on Deep-Inelastic Scattering and
Related Subjects, \\
Michigan State University, USA, 27-31 March 2023 \\
         \includegraphics[width=9cm]{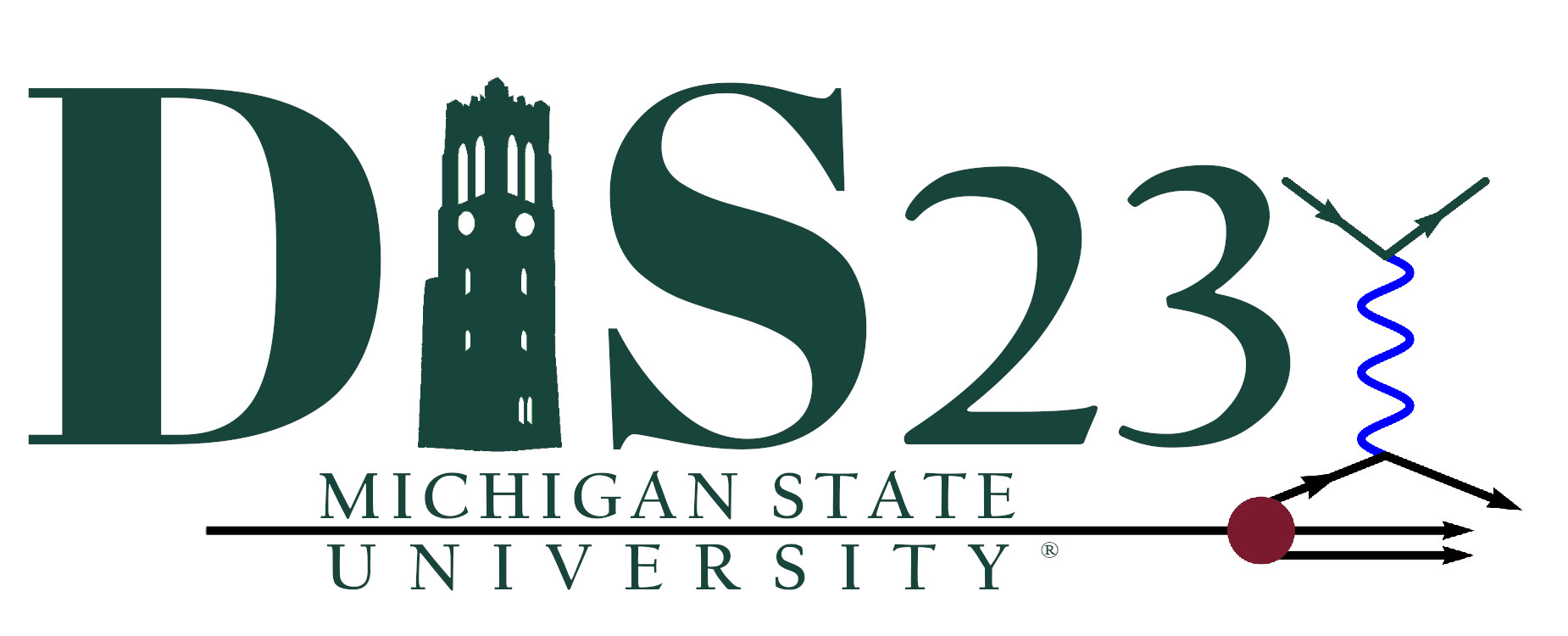}
\end{Presented}
\vfill
\end{titlepage}

\section{Introduction}
An interesting discovery by Fermilab was published in 1976. They observed that $\Lambda$ hyperons produced in $p+$Be collisions with a 300~GeV proton beam are polarized \cite{ref-first_paper}. This observation is surprising because neither the proton beam nor the beryllium target was polarized. As a result, experimentalists and theorists all around the world started investigating this phenomenon.

\begin{figure}[ht]
  \centering
  \includegraphics[width=0.5\textwidth]{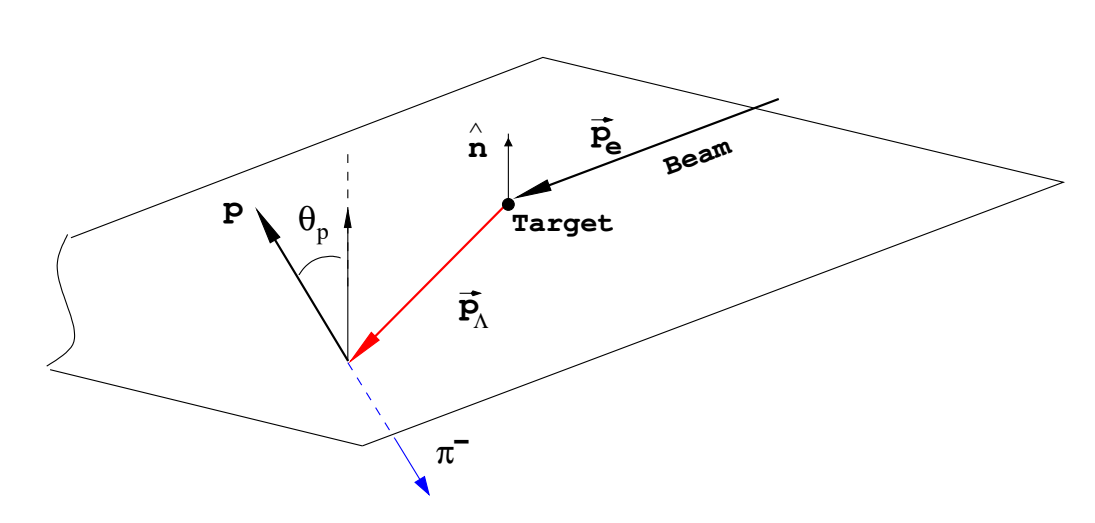}
  \caption{ Cartoon illustrating production plane determination. The plane is defined by momentum of the beam ($\vec{p_\mathrm{e}}$) and momentum of $\Lambda$ ($\vec{p_\Lambda}$). Normal vector $\hat{n}$ to the plane is then given by $\hat{n} = \vec{p_\mathrm{e}} \times \vec{p_\Lambda}$. The $\Lambda$ polarization is then quantified by measuring the angle $\theta_\mathrm{p}$ between the momentum of proton in the rest frame of its mother and $\hat{n}$. Taken from Ref. \cite{ref-HERMES}.}
  \label{fig_prod_plane}
\end{figure}

The $\Lambda$ hyperon polarization can be measured through reconstruction of a hadronic decay channel $\Lambda^0 \rightarrow p\pi^-$ (and charge conjugate) and subsequent measurement of the angle ($\theta_\mathrm{p}$, or $\theta^\star$) between the decay proton momentum in the $\Lambda$ rest frame ($p$) and a normal vector to the production plane ($\hat{n}$). This decay channel is selected, because the proton is preferentially emitted in the direction of the $\Lambda$ polarization in the $\Lambda$ rest frame. A cartoon illustrating the production plane determination, using the variables defined above, is shown in Fig. \ref{fig_prod_plane}.

The polarization ($P_\Lambda$) is then extracted from the angular distribution of the protons according to formula

\begin{equation}\label{eq_pol_standard}
    \frac{\mathrm{d}N}{\mathrm{d}\cos(\theta^\star)} = 1 + \alpha P_\Lambda \cos(\theta^\star),
\end{equation}

\noindent where $\alpha$ is the weak decay constant of the $\Lambda$ hyperon. This method was used in the first $\Lambda$ hyperon polarization measurement from Ref. \cite{ref-first_paper} and other measurements that followed. It is also possible to measure the $\Lambda$ hyperon polarization with respect to a different reference direction, e.g. a jet axis, or polarization of the beam for measurements with polarized beams. A brief overview of experimental results using these traditional methods is provided in Sec. \ref{sec_results}.

Section \ref{sec_correlations} provides a description of a new method for $\Lambda$ hyperon polarization measurement, which relies on the determination of $\Lambda$ hyperon pair spin-spin correlation. In addition, the section shows first steps of analysis utilizing this method in $p+p$ collisions at $\sqrt{s} = 200$~GeV by the STAR experiment.


\section{Overview of $\Lambda$ hyperon polarization results}\label{sec_results}
\begin{figure}[ht]
  \centering
  \includegraphics[width=0.55\textwidth]{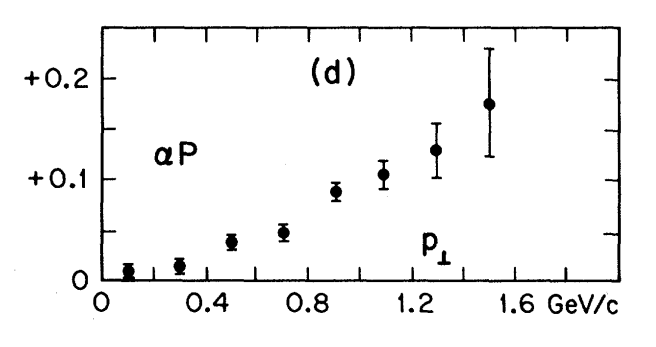}
  \caption{ The first observation of $\Lambda$ hyperon polarization as a function of transverse momentum, measured at Fermilab in $p+$Be collisions with 300 GeV proton beam. Taken from Ref \cite{ref-first_paper}.}
  \label{fig_first_paper}
\end{figure}

One of the results presented in the first $\Lambda$ hyperon polarization paper is shown in Fig. \ref{fig_first_paper}. The polarization $\alpha P_\Lambda$ of the $\Lambda$ hyperons produced in $p+$Be collisions with 300~GeV proton beam on a Be target, rises with the $\Lambda$ transverse momentum ($p_\mathrm{T}$).

\begin{figure}[ht]
  \centering
  \includegraphics[width=0.5\textwidth]{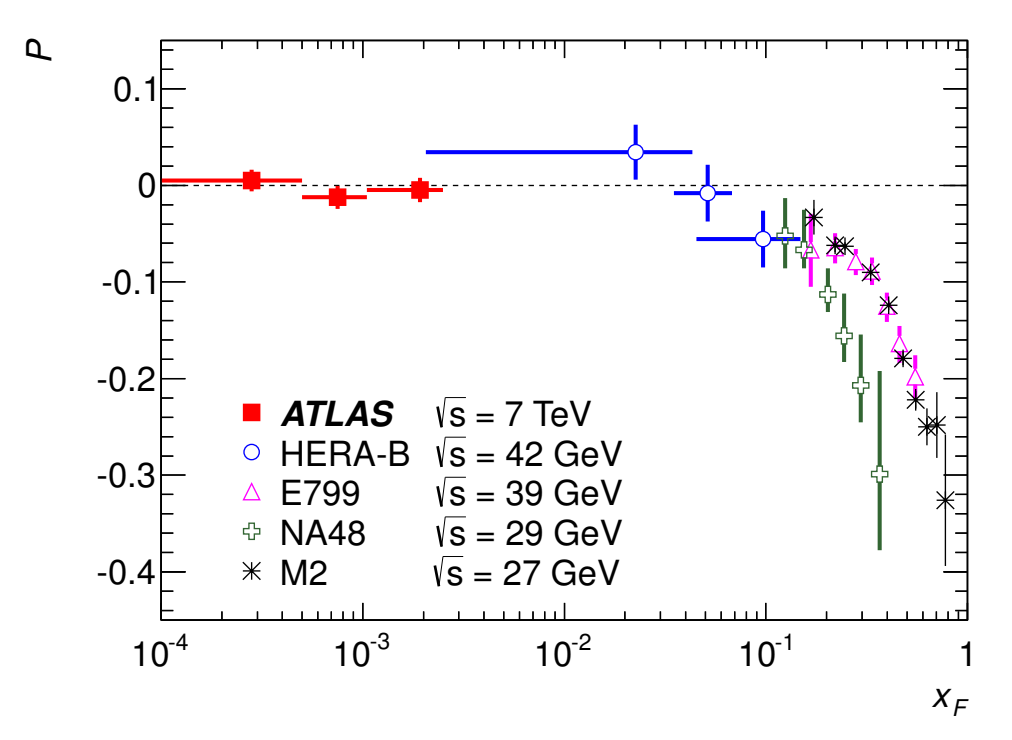}
  \caption{ Comparison of the $\Lambda$ hyperon polarization as a function of $x_\mathrm{F}$ measured in variety of collision systems and collision energies. Taken from Ref. \cite{ref-ATLAS}.}
  \label{fig_ATLAS}
\end{figure}

In pursuit of an explanation of the origin of $\Lambda$ hyperon polarization, a number of independent investigations were performed over a wide range of collision systems and energies. A selection of such results is shown in Fig. \ref{fig_ATLAS}. In this case, the polarization $P_\Lambda$ is plotted as a function of $x_\mathrm{F} = p_\mathrm{z}^\Lambda/p_\mathrm{beam}$. The key observation here is that the polarization appears to depend primarily on the $x_\mathrm{F}$ and not on the collision type or energy.

\begin{figure}[ht]
  \centering
  \includegraphics[width=0.5\textwidth]{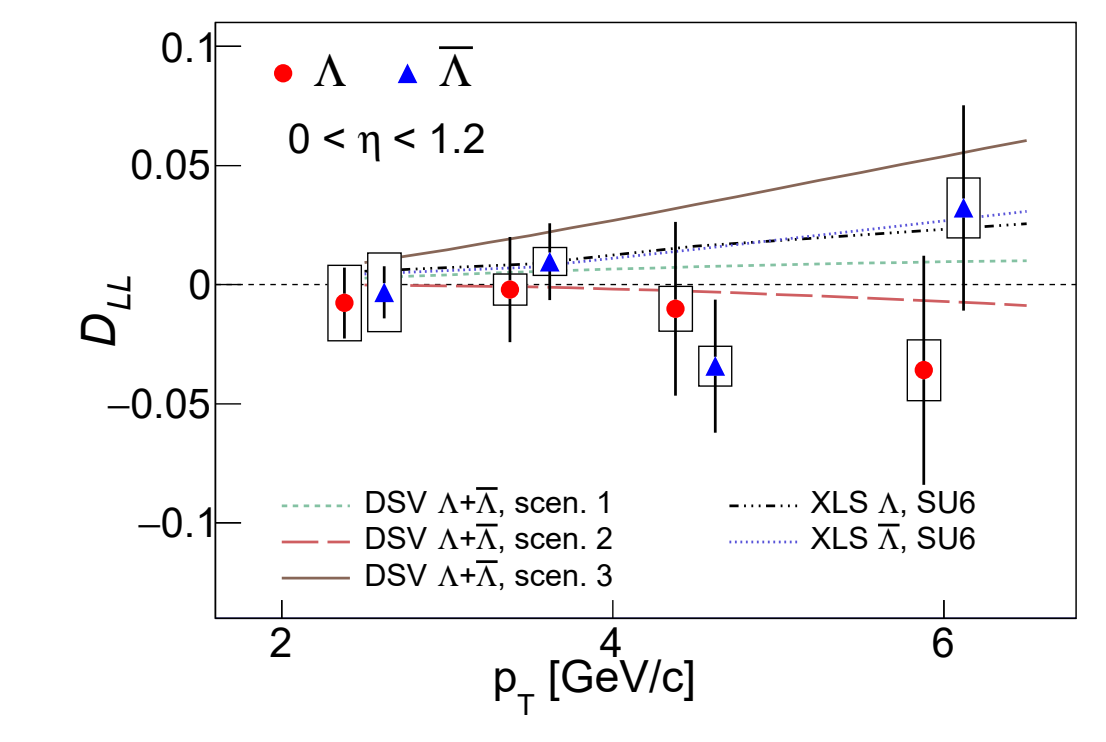}
  \caption{ Longitudinal spin transfer $D_\mathrm{LL}$ of $\Lambda$ (red circles) and $\bar{\Lambda}$ (blue circles) hyperons at positive pseudorapidity ($0 < \eta < 1.2$) measured in $\vec{p}+\vec{p}$ collisions at $\sqrt{s} = 200$~GeV by the STAR experiment. Taken from Ref \cite{ref-STAR_transfer}.}
  \label{fig_STAR_spin_transfer}
\end{figure}

All presented results are from collisions of unpolarized particles. It is also important to investigate, if the polarization of the produced $\Lambda$ hyperons is correlated with polarization of the beams, for example, in polarized $\vec{p}+\vec{p}$ collisions at the STAR experiment. An example of such measurement is presented in Fig. \ref{fig_STAR_spin_transfer} which shows the longitudinal spin transfer $D_\mathrm{LL}$ of $\Lambda$ and $\bar{\Lambda}$ hyperons at positive pseudorapidity ($0 < \eta < 1.2$) measured in $\vec{p}+\vec{p}$ collisions at $\sqrt{s} = 200$~GeV. No significant longitudinal polarization of $\Lambda$ hyperons is observed at STAR, which suggests that the beam polarization does not play a significant role for the polarization of the $\Lambda$ hyperon at RHIC energies within the studied kinematic range\footnote{The $x_\mathrm{F}$ in this $\eta$ region at RHIC is going to be rather small which will likely lead to small $\Lambda$ polarization, as seen in Fig. \ref{fig_ATLAS}. It is not possible to make direct comparison to Fig. \ref{fig_ATLAS} due to polarization of the beams and also a different observable of the measurement.}.

The examples above are a small selection of all experimental efforts over the last 50 years in attempt to explain the origin of $\Lambda$ hyperon polarization. Unfortunately, none of the measurements, or the theoretical models, can provide a definitive answer on where the $\Lambda$ hyperon polarization is generated. In the following section, we investigate a possibility to improve our knowledge of the phenomenon by measurement of $\Lambda$ hyperon pair spin-spin correlations.

\FloatBarrier
\section{$\Lambda$ hyperon spin-spin correlations}\label{sec_correlations}
In general, most experimental techniques for measurement of $\Lambda$ hyperon polarization are based on the same general idea. It is the measurement of an angle, usually denoted $\theta^\star$, between a reference direction and momentum of the decay proton in the hyperon rest frame. The reference direction can be chosen based on specific physics considerations. As shown above, the first possibility is to use the production plane. Other common alternatives are, e.g., the polarization of the beam, or a jet axis, in case the $\Lambda$ hyperon is part of a jet.

Another possibility is to look for events with two or more $\Lambda$ or $\bar{\Lambda}$ hyperons and measure the angle $\theta^\star_{12}$ between the decay (anti-)protons, both boosted into the rest frame of their mother particles. Since these (anti-) protons are preferentially emitted in the direction of their mother's polarization, such measurement gives access to $\Lambda\bar{\Lambda}$, $\Lambda\Lambda$, and $\bar{\Lambda}\bar{\Lambda}$ pair spin-spin correlations \cite{ref-Tornqvist,ref-Gong}. For this method, it is possible to use the following formula (see also Ref. \cite{ref-Gong}):

\begin{equation}\label{eq_pol_new}
    \frac{\mathrm{d}N}{\mathrm{d}\cos(\theta^\star_{12})} = 1 + \alpha_1\alpha_2 P_{\Lambda_1\Lambda_2} \cos(\theta^\star_{12}),
\end{equation}

\noindent where $\alpha_1$ and $\alpha_2$ are weak decay constants of $\Lambda$ hyperons in the pair and $P_{\Lambda_1\Lambda_2}$ is the polarization of the pair. $\Lambda_1$ and $\Lambda_2$ can both be either $\Lambda$ or $\bar{\Lambda}$.

One of the key advantages of this approach is that it should be able to identify if the polarization comes from initial stage effects, such as spin-spin correlation of the initial stage partons, or if it is a final state effect originating from e.g. hadronization. The initial state correlation should be seen in data as spin-spin correlation of the $\Lambda\bar{\Lambda}$ pairs, as those likely originate from a single $s\bar{s}$ quark pair produced in the hard partonic scattering. At the same time, no strong correlation is expected for $\Lambda\Lambda$ and $\bar{\Lambda}\bar{\Lambda}$ pairs, as those cannot originate from a single $s\bar{s}$ quark pair.

\begin{figure}[ht]
  \centering
  \begin{subfigure}[c]{0.48\textwidth}
    \includegraphics[width=\textwidth]{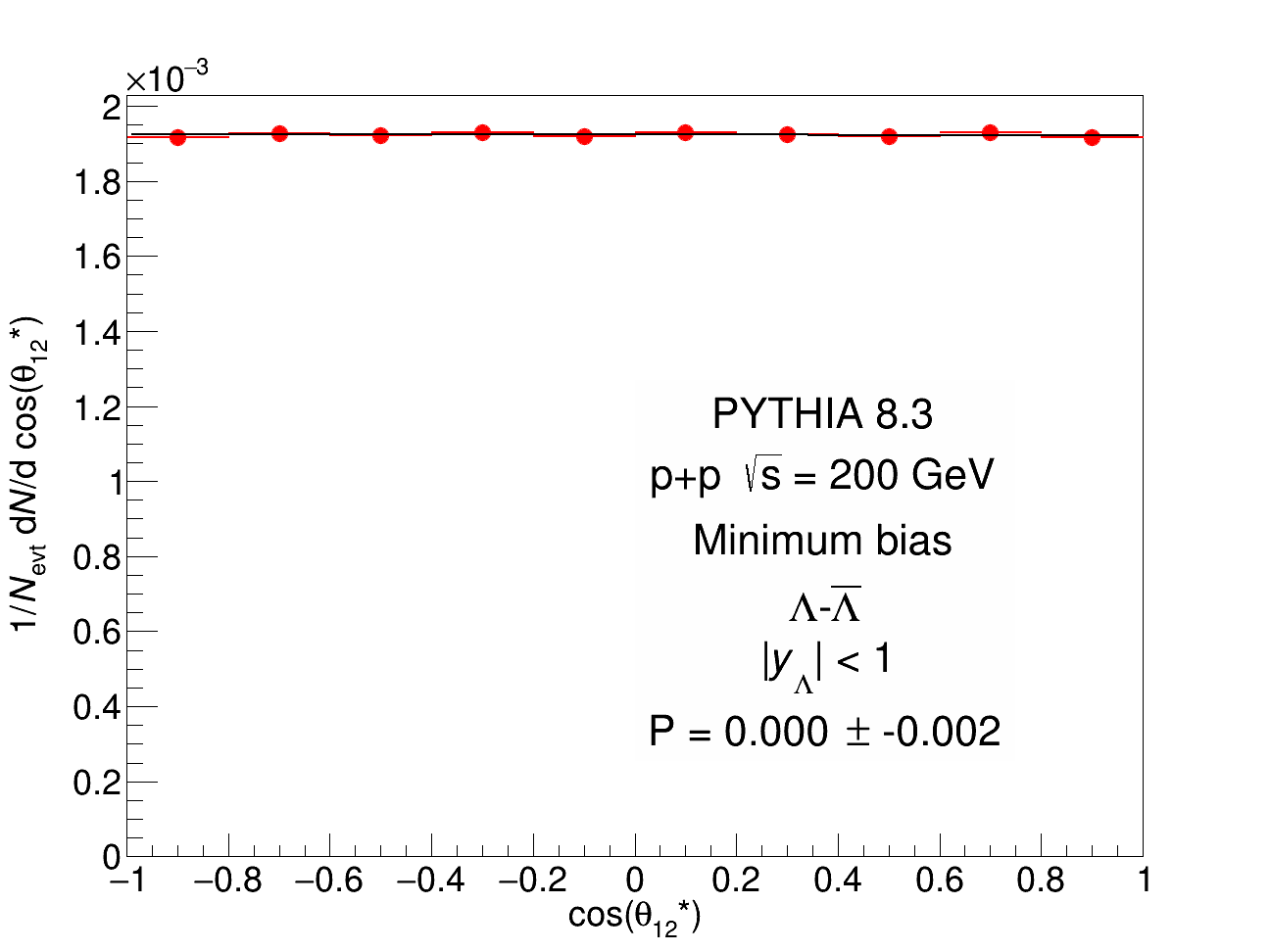}
  \end{subfigure}
  \begin{subfigure}[c]{0.48\textwidth}
    \includegraphics[width=\textwidth]{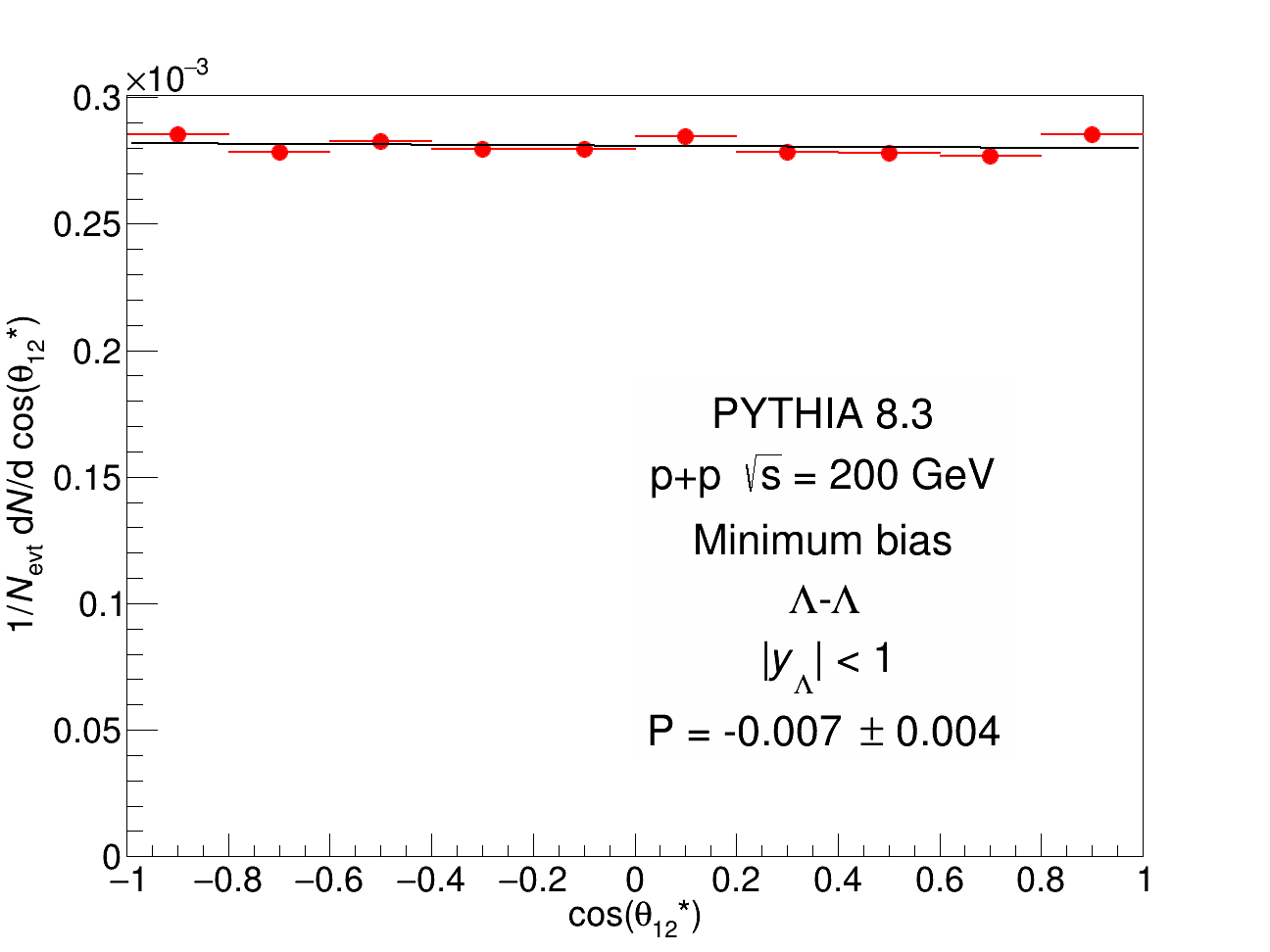}
  \end{subfigure}
  \caption{ $1/N_\mathrm{evt}\textrm{d}N/\textrm{d}\cos(\theta^\star_{12})$ as a function of $\cos(\theta^\star_{12})$ for $\Lambda\bar{\Lambda}$ (left) and  $\Lambda\Lambda$ (right) pairs from PYTHIA 8.3 $p+p$ collisions at $\sqrt{s} = 200$~GeV. The simulation is fitted with a linear function which is used to extract the polarization $P$ according to equation (\ref{eq_pol_new}).}
  \label{fig_STAR_spin_transfer_PYTHIA}
\end{figure}

In order to investigate the $\Lambda$ hyperon pair spin-spin correlations in $p+p$ collisions at $\sqrt{s} = 200$~GeV at STAR, it is important to verify that there are no other known mechanisms to generate non-zero $P_{\Lambda_1\Lambda_2}$. This was done with PYTHIA 8.3 simulation of $p+p$ collisions at $\sqrt{s} = 200$~GeV. The extracted $1/N_\mathrm{evt}\mathrm{d}N/\mathrm{d}\cos(\theta^\star_{12})$ distributions as a function of $\cos(\theta^\star_{12})$ for $\Lambda\bar{\Lambda}$ and $\Lambda\Lambda$ pairs are shown in Fig. \ref{fig_STAR_spin_transfer_PYTHIA}. The distributions are fitted with a linear function which is used to extract the value of $P_{\Lambda_1\Lambda_2}$ using equation (\ref{eq_pol_new}). For both combinations, the polarization is zero, meaning that pure PYTHIA does not predict any $\Lambda$ hyperon spin-spin correlations at mid-rapidity in $p+p$ collisions at $\sqrt{s} = 200$~GeV.

The extraction of the $\mathrm{d}N/\mathrm{d}\cos(\theta^\star_{12})$ distributions from the data starts with selection of $\Lambda$ and $\bar{\Lambda}$ hyperon candidates. This is done by pairing protons and pions reconstructed and identified with the STAR Time Projection Chamber. Each $\Lambda$ candidate then corresponds to one $p\pi$ pair. Events with two or more $\Lambda$ candidates were considered for further analysis. For events which contain at least two such pairs, a 2D distribution is filled where one axis is the invariant mass of one of the $p\pi$ pairs and the second axis is the invariant mass of the second $p\pi$ pair.

\begin{figure}[ht]
  \centering
  \begin{subfigure}[c]{0.48\textwidth}
    \includegraphics[width=\textwidth]{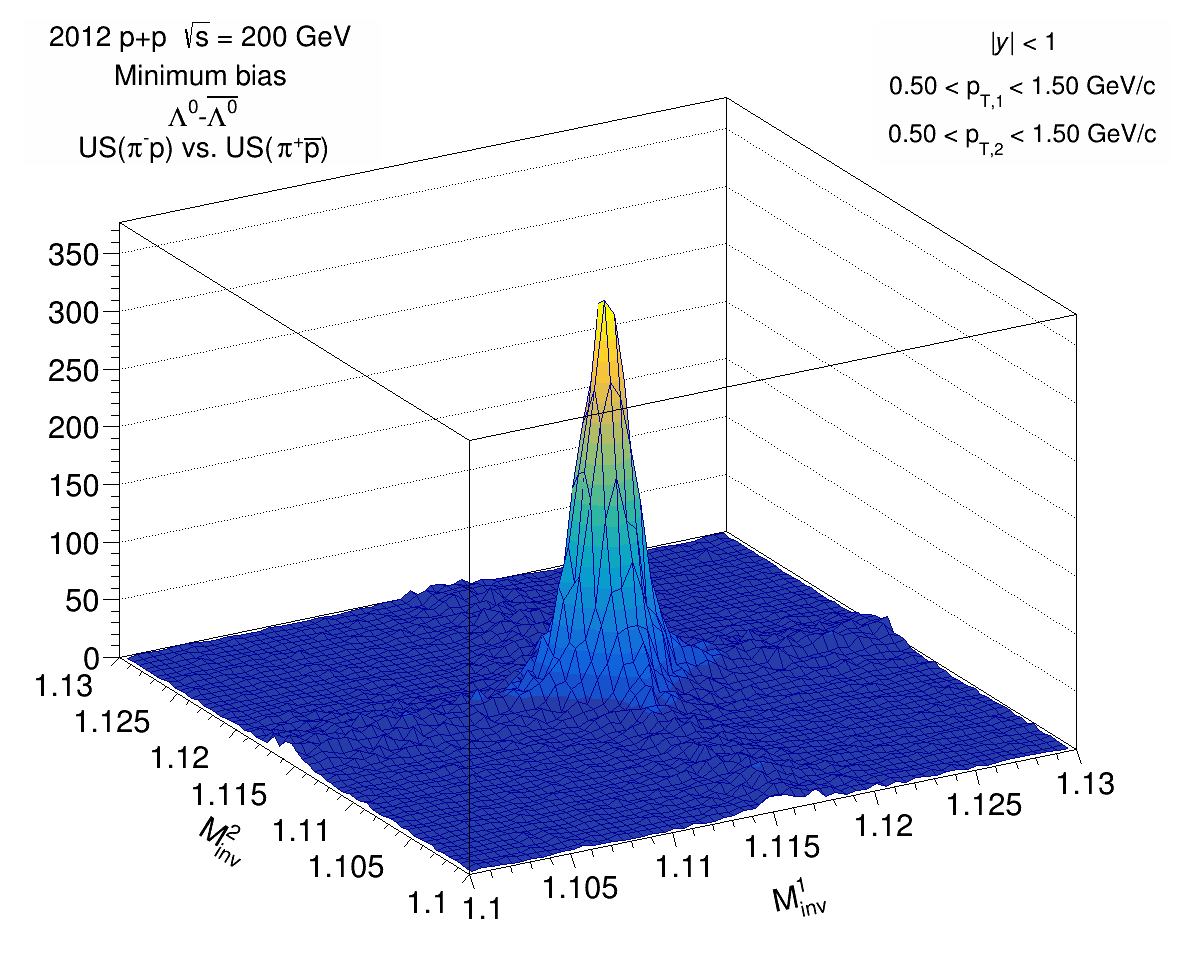}
  \end{subfigure}
  \begin{subfigure}[c]{0.48\textwidth}
    \includegraphics[width=\textwidth]{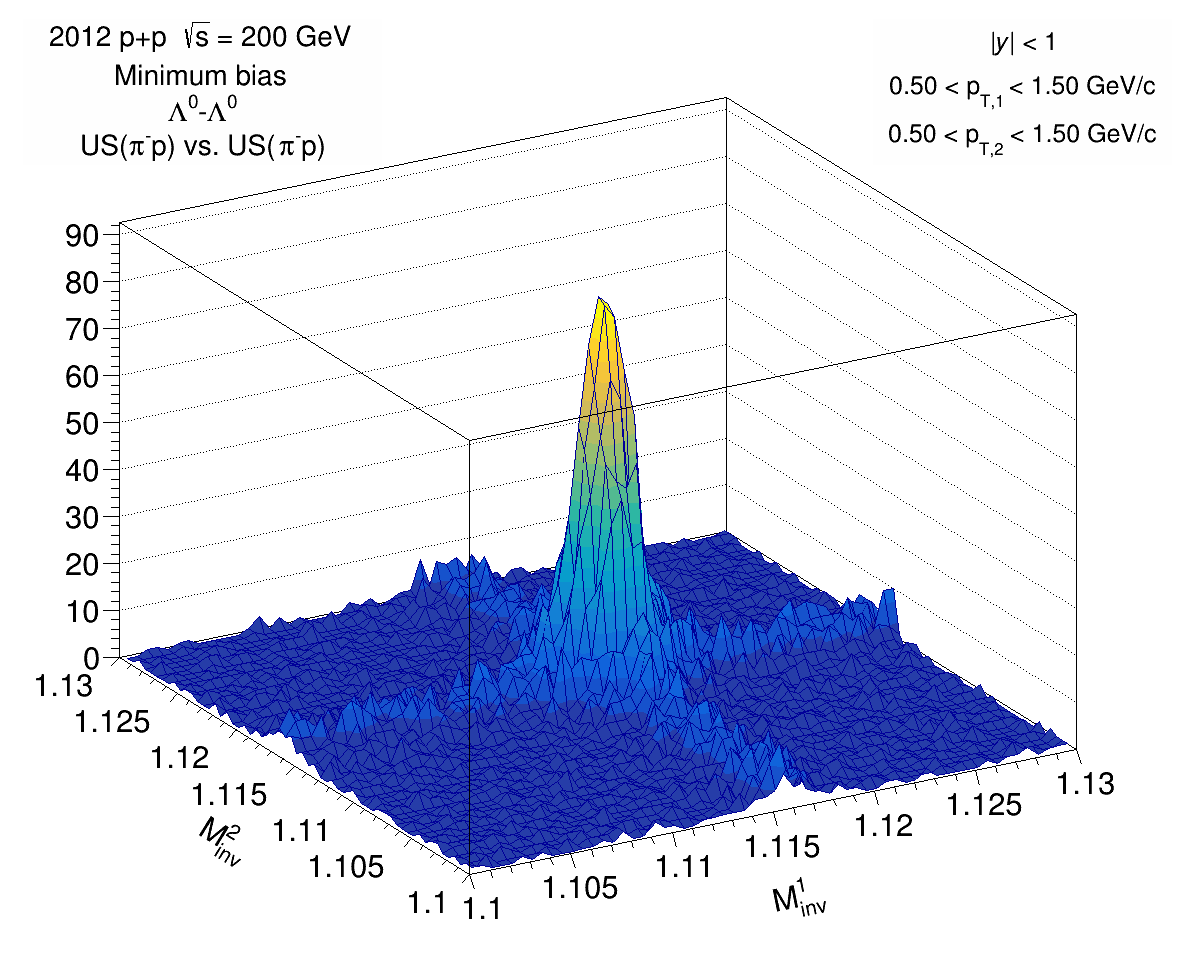}
  \end{subfigure}
  \caption{ Two-dimensional invariant mass distributions of unlike-sign $p\pi$ pairs matched with unlike-sign $p\pi$ pairs. Left panel shows $p\pi$ pair charge combinations corresponding to $\Lambda\bar{\Lambda}$ pair candidates, the right panel shows the $\Lambda\Lambda$ pair candidates.}
  \label{fig_STAR_inv_mass}
\end{figure}

This is done for two combinations of the $p\pi$ pairs: an unlike-sign (US) $p\pi$ pair matched with a different US pair from the same event. This distribution contains both the signal and the combinatorial background. The background can be estimated using US pairs matched to the like-sign (LS) $p\pi$ pairs. The US-LS distribution is then subtracted from US-US distribution and subsequently fitted with a 2D Gaussian function to determine the $\Lambda$ candidate invariant mass peak mean and width. The signal region is defined as the mean $\pm3\sigma$, where both the mean and $\sigma$ are taken from the fit. Example of two of the 2D invariant mass distributions is shown in Fig. \ref{fig_STAR_inv_mass}.

This procedure is done separately for $\Lambda\bar{\Lambda}$, $\Lambda\Lambda$, and $\bar{\Lambda}\bar{\Lambda}$ candidate pairs in $p+p$ collisions at $\sqrt{s} = 200$~GeV measured by STAR in 2012. The extracted numbers of signal and background pairs for each of the three possible combinations is shown in Tab. \ref{tab_pair_stats}. The number of candidate pairs is going to provide sufficient statistics to perform this measurement using the 2012 $p+p$ collisions data-set.

\begin{table}[ht]
\centering
    \begin{tabular}{ccccccc}
    \toprule
    \multirow{2}{*}{$p_\mathrm{T}$ [GeV/$c$]} & \multicolumn{2}{c}{$\Lambda\bar{\Lambda}$} &  \multicolumn{2}{c}{$\Lambda\Lambda$} &   \multicolumn{2}{c}{$\bar{\Lambda}\bar{\Lambda}$} \\
    \cmidrule{2-7}
    & $N_\mathrm{sig}$ & $N_\mathrm{bkg}$ & $N_\mathrm{sig}$ & $N_\mathrm{bkg}$ & $N_\mathrm{sig}$ & $N_\mathrm{bkg}$ \\
    \midrule
    $\Lambda_1$ $0.5 - 5.0$ & \multirow{2}{*}{26767} & \multirow{2}{*}{6929} & \multirow{2}{*}{7090} & \multirow{2}{*}{3593} & \multirow{2}{*}{4670} & \multirow{2}{*}{2495} \\
    $\Lambda_2$ $0.5 - 5.0$ & & & & & & \\
    \bottomrule
    \end{tabular}
\caption{Number of $\Lambda\bar{\Lambda}$, $\Lambda\Lambda$, and $\bar{\Lambda}\bar{\Lambda}$ signal candidate and background pairs in $p+p$ collisions at $\sqrt{s} = 200$~GeV measured by STAR in 2012. }
\label{tab_pair_stats}
\end{table}

\section{Summary}
The $\Lambda$ hyperon polarization puzzle is one of the main unresolved mysteries of the experimental high energy particle physics. The polarization has been observed in several different collision systems at various energies. The magnitude of the polarization appears to depend primarily on $x_\mathrm{F}$ and not much on the specific collision energy. Despite enormous experimental and theoretical efforts to explain the $\Lambda$ hyperon polarization, no conclusive answer was found. In order to improve the knowledge in this field, a new method was developed which relies on measurement of $\Lambda$ hyperon pair spin-spin correlations. Any non-zero signal measured in $p+p$ collisions at $\sqrt{s} = 200$~GeV at STAR would provide more insight into $\Lambda$ hyperon polarization, as a simulation using PYTHIA 8.3 predicts no spin-spin correlation signal. The number of extracted $\Lambda\bar{\Lambda}$, $\Lambda\Lambda$, and $\bar{\Lambda}\bar{\Lambda}$ candidate pairs extracted from the aforementioned STAR data-set is sufficient to perform this type of measurement and thus is going to provide important additional insight into $\Lambda$ hyperon polarization physics in $p+p$ collisions at RHIC energies.

\end{document}